\documentclass[prb,aps,epsf]{revtex4}
\usepackage{amsmath}
\input epsf
\begin{document}
\draft 
\title{Projectiles, pendula, and special relativity}
\author{Richard H. Price}
\affiliation{
Department of Phyiscs and Astronomy
and Center for Gravitational Wave Astronomy,
The University of Texas as Brownsville, Brownsville,
Texas, 78520
}

\begin{abstract}
The kind of flat-earth gravity used in introductory physics appears
in an accelerated reference system in special relativity. From this
viewpoint, we work out the special relativistic description of a
ballistic projectile and a simple pendulum, two examples of simple
motion driven by earth-surface gravity. The analysis uses only the
basic mathematical tools of special relativity typical of a
first-year university course.
\end{abstract}

\maketitle

\section{Introduction}\label{sec:intro} 

Students often see special relativity as having an inherent
elegance, but see no overlap with the problems they
study in an introductory physics course. To tie the two student
experiences closer together, this paper presents a relativistic
treatment of two of the simpler problems studied in elementary
mechanics courses: ballistic motion and the simple pendulum. 

These familiar examples share an important feature: in both
problems, the driving force is the weight of gravity. This feature
opens the possibility of treating these problems relativistically
by elementary means. This possibility arises due to the equivalence
principle. All objects have the same acceleration in a
gravitational field; a rock and a feather fall at the same rate in
a gravitational field. Consequently, objects in a freely falling
frame (like a falling elevator) have no apparent weight. 

In relativistic gravitation theories, like Einstein's general
relativity, the interpretation of the equivalence principle goes in
the opposite direction. In these theories it is the freely falling
frame that is the inertial frame. In a frame that is not freely
falling, an apparent weight force arises. But it is a pseudoforce, a
fictitious force like the centrifugal force, and is an artifact of
the noninertial reference frame.\cite{taylorwheeler,mtw} Thus a
downward apparent weight force is perceived on the surface of the
earth because the surface of the earth is accelerating upward at
9.81\,m/s$^2$ with respect to a freely falling frame.

Here we exploit this point of view. To study ballistic motion and
pendula, we do not need a complete relativistic theory of gravity;
we need only to account for the weight force
(or pseudoforce). We work completely in gravity-free special
relativity, and we introduce weight by working in an
upward accelerating special relativistic reference frame.

This article is intended for students who have had only the beginnings
of relativity. All that is required is the simple (one-dimensional)
Lorentz transform, the concept of proper time along a worldline, the
acceleration four-vector, and the fact that its components in
different reference frames are related by the Lorentz
transformation. For more advanced students, the results may still be
of interest, and are more easily derived with techniques like
covariant differentiation. For such students, a more compact
presentation, using more advanced techniques, is given in
Appendix~\ref{sec:appendixB}.                        
In order to give a more
accessible description, some details are relegated to the endnotes.

\section{Earthlike frame}\label{sec:earthlike} 
We may take the relativistic point of view in introductory Newtonian
physics by invoking a freely falling $t,x,y,z$ reference frame near
the surface of a flat earth, and by considering this reference frame
to be an inertial frame. In this inertial frame, the $x,y,z$ spatial
coordinates constitute a Cartesian spatial grid, and $t$ is the
universal time of Newtonian physics.
\begin{figure}[h]
\begin{center}
{\epsfxsize=260pt \epsfbox{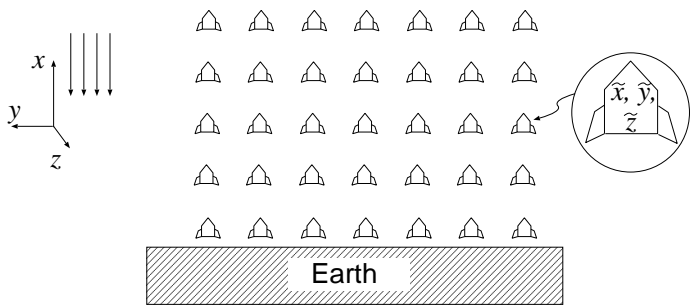}}
\end{center}
\caption{\label{fig:rocksts} The earth-like
$\widetilde{x},\widetilde{y},\widetilde{z}$ spatial coordinate
system is kept stationary with respect to the earth by the thrust
provided by the rocket engines. 
The inertial $x,y,z$ coordinate system is freely falling.}
\end{figure}
 
We then imagine a swarm of tiny rocket ships accelerating upward
with respect to the inertial frame. At $t=0$
each of the rockets is (momentarily) at rest. We assign labels
$\widetilde{x},\widetilde{y},\widetilde{z}$ to the rockets, such
that at $t=0$ we have $\widetilde{x}=x$, $\widetilde{y}=y$ and
$\widetilde{z}=z$. Because the rockets move in the $x$ direction,
they maintain $\widetilde{y}=y$ and $\widetilde{z}=z$, but $x$ for
each rocket is a function of time and it is this time dependence
that make our rocket-borne reference frame an accelerated,         
and therefore                                                              
noninertial, frame with a weight
force like that of the earth's surface.

We clearly want to have our rocket-borne reference frame to be
accelerating upward. In Newtonian mechanics this acceleration would
be done by choosing
\begin{equation}\label{Newtxoft} 
x=\widetilde{x}+{\textstyle\frac{1}{2}}gt^2\,.
\end{equation}
With this choice each rocket moves in the same way and the spatial
$\widetilde{x},\widetilde{y},\widetilde{z}$ grid is always a
Cartesian system for measuring distances. Newton's second law can be
used in this noninertial system if every mass element $m$ is taken
to have a gravitational (pseudo)force $mg$ acting on it in the
negative
$\widetilde{x}$ direction. We can do mechanics either in the freely
falling inertial frame with no gravity or in the noninertial frame
with gravity.

In special relativity the equivalent construction has some new
subtleties. Our primary reference frame $t,x,y,z$ is now a Minkowski
coordinate system, with no gravity. As in the Newtonian case, we
again invoke the swarm of rockets that are momentarily at rest in
the
$t,x,y,z$ frame at $t=0$. Again we assign labels to the rockets such
that $\widetilde{y}=y$ and $\widetilde{z}=z$ for all $t$. We must
now choose how the rockets are to accelerate, that is, we must
specify
$x(t)$ for each rocket.

It turns out that the $x(t)$ in Eq.~(\ref{Newtxoft}) is not ideal in
special relativity. The reference frame created by that $x(t)$ would
have undesirable features. For example, a rocketeer might want to
measure the distance from her rocket to another nearby rocket. This
measurement would be done in her momentarily comoving frame.  The
distance measured in this way will change in time. The rocket frame,
then, would not be an unchanging frame like the reference frame used in
introductory Newtonian mechanics. 

The choice of $x(t)$ that is close to ideal turns out to
be\cite{rindlerrefs} 
\begin{equation}\label{x2minusc2t2} 
x^2-c^2
t^2
=\kappa^2
\,,
\end{equation} where $\kappa$ is a constant. The reason for
favoring this choice is not at all obvious, but at least one of its
features is comforting. For $ct\ll\kappa$, Eq.~(\ref{x2minusc2t2})
becomes
\begin{equation}
x\approx\kappa+\frac{1}{2}\;\frac{c^2}{\kappa}t^2\,.
\end{equation}
Thus, when a rocket is moving at nonrelativistic velocity
($dx/dt\approx c^2t/\kappa\ll c$), this relativistic choice of
$x(t)$ takes the Newtonian form in Eq.~(\ref{Newtxoft}) if we take 
the acceleration to be to be $c^2/\kappa$. 

The constant $\kappa$ in Eq.~(\ref{x2minusc2t2}) can be different
for each rocket, as in Eq.~(\ref{Newtxoft}), so that this constant
can be used to assign an $\widetilde{x}$ coordinate to each rocket.
As we shall demonstrate, it is best to do this by
choosing\cite{different_g}
\begin{equation}\label{SRTxoft} 
x^2-c^2t^2=(\widetilde{x}+c^2/g)^2 .
\end{equation}
The meaning of $\widetilde{x}$ is potentially confusing. It is a
constant along the world line of any particular rocket. But we
will also use it as a spatial label in the
$\widetilde{x}$, $\widetilde{y}$, $\widetilde{z}$ system. A particle
--- like a ballistic projectile or a pendulum bob --- moving from
one rocket location to another would have a time varying value of 
$\widetilde{x}$, and it is meaningful to consider 
$\widetilde{x}(t)$ for such a particle.

The $\widetilde{x},\widetilde{y},\widetilde{z}$ system will be our
our earth-like system. It is to be considered a spatial reference
frame only and is not part of a Minkowski system. To help avoid
confusion we will not (except in Appendix~\ref{sec:appendixB})
endow this reference frame with an associated time coordinate.
Rather, we will discuss the dynamics of particles (ballistic
projectiles and pendulum bobs) with the proper time $\tau$ for
those particles, the time measured by clocks carried on the
particles.

\begin{figure}[h]
 \begin{center}
{\epsfxsize=170pt \epsfbox{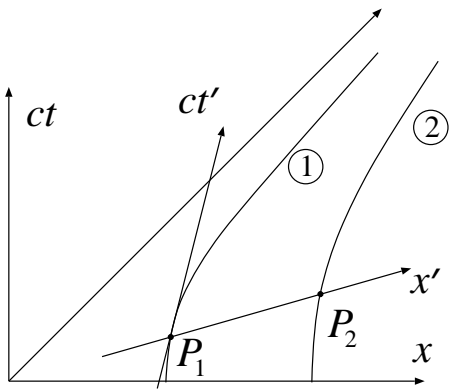}}
 \end{center}
\caption{\label{fig:spacetime} 
In a $t,x$ spacetime diagram, the worldlines
of rockets 1 and 2 are shown, along with the coordinate axes
of the $t',x'$ Minkowski reference frame, the frame comoving
with rocket 1 at event $P_1$.
}
\end{figure}

The real justification for Eq.~(\ref{SRTxoft}) is that with this
choice, the rocket-borne $\widetilde{x},\widetilde{y},\widetilde{z}$
reference frame has three important properties that qualify it as an
earth-like spatial reference frame. These properties are best
understood with a spacetime diagram like that in
Fig.~\ref{fig:spacetime}. The diagram shows worldlines for two
arbitrary rockets, labeled 1 and 2. According to Eq.~(\ref{SRTxoft})
these worldlines are hyperbolae asymptotic to $x=ct$. In this
diagram
$P_1$ (coordinates $t_1,x_1$) is an event on the worldline of rocket
1, and $t',x',y',z'$ is a Minkowski coordinate system
instantaneously comoving with rocket 1 at event $P_1$. Event $P_2$
(coordinates
$t_2,x_2$) is the event on the worldline of rocket 2 that is
simultaneous, in the instantaneously comoving frame, with $P_1$.
That is, events $P_1$ and $P_2$ have the same value of $t'$, and
are simultaneous as seen in the reference frame of $P_1$.

The three special properties of the earth-like system include the
following. (i) In a frame that is instantaneously comoving with one
rocket, all rockets are instantaneously at rest. (In other words,
the velocity
$dx/dt$ of rocket 2 at $P_2$ is the same as that of rocket 1 at
$P_1$.) (ii) The acceleration of each rocket is a constant in
time. (iii) The distance between any two rockets, as
measured in an instantaneously comoving frame, is time independent.
These properties, proved in Appendix~\ref{sec:appendixA}, establish
that the rocket-borne $\widetilde{x},\widetilde{y},\widetilde{z}$
system is a ``rigid'' framework for spatial measurements. Although
it is not part of an inertial reference frame, it is, in a
sense, the same at all times.

\section{Ballistic trajectories}\label{sec:ballistic}

A ballistic projectile has only the weight force acting on it. This
means that its worldline will be straight in a
Minkowski coordinate system.\cite{lapidus} With little loss in
generality we choose the projectile to be moving in the $x,y$
plane, with $x$ having the fixed value $x=c^2/g$, so that the
projectile starts off with
$\widetilde{x}=0$ at $t=0$. In the freely falling $t,x,y,z$ frame,
the projectile is moving only in the $y$ direction, and we
specify its motion by $y=vt\equiv c\beta t$. A moment of proper time
$d\tau$ and of coordinate time $dt$ are related, as usual, by
$d\tau=\sqrt{(1-\beta^2)}\,dt\equiv dt/\gamma$.

We choose $\tau$ to be zero when $t$ is zero, and the complete description 
of the projectile motion in the freely falling frame becomes
\begin{equation}
t=\gamma\tau,\qquad x={c^2}/{g}, \qquad y=c\beta\gamma\tau.
\end{equation}
With Eq.~(\ref{SRTxoft}), the description in the earth-based 
frame immediately follows:
\begin{equation}
\label{traj} 
\widetilde{x}=-{c^2}/{g}+\sqrt{({c^2}/{g})^2
-c^2\gamma^2\tau^2}, \qquad \widetilde{y}=c\beta\gamma\tau\,,
\end{equation}
and we see that the 
trajectory has the shape of an ellipse
\begin{equation}
(\widetilde{x}+{c^2}/{g})^2
+\widetilde{y}^2/\beta^2=({c^2}/{g})^2 .
\end{equation}

\begin{figure}[h]
 \begin{center}
{\epsfxsize=110pt \epsfbox{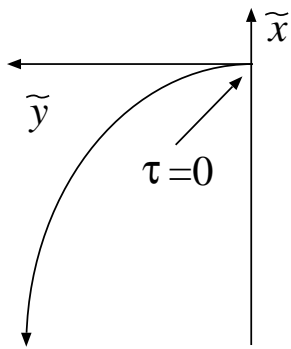}}
 \end{center}
\caption{\label{fig:ellipse} 
Ballistic trajectory for $v=4c/5$, starting at $\tau=0$ and ending at $\tau
=(3/5)(c/g)$\,.
}
\end{figure}

If $v\ll c$ and $\tau\ll c/g\approx 1$\,year, then Eq.~(\ref{traj})
reduces to
\begin{equation}
\label{eq:results}
y\approx v\tau, \qquad\ 
\widetilde{x}\approx-\frac{1}{2}\,g\tau^2
\approx-\frac{g}{2v^2} \widetilde{y}^2 .
\end{equation}
It is reassuring that Eq.~(\ref{traj}) has the familiar
nonrelativistic limit, but it is more interesting in its fully
relativistic form. 
Figure~\ref{fig:ellipse}
shows the elliptical trajectory for the case $v=4c/5$. This ellipse
differs noticeably from a parabola. It is of interest that the
trajectory ends at the point $\widetilde{x}=-c^2/g$\,,\ 
$\widetilde{y}=(4/5)c^2/g$, corresponding to the proper time
$\tau=(3/5)c/g$. This sudden end is not an indication of a dramatic
physical event. Rather, it signals the limit of the ability of the
$\widetilde{x}$ $,\widetilde{y}$
$,\widetilde{z}$ coordinates to cover the
spacetime.\cite{explainrindler}

\section{Simple pendulum}\label{sec:pendulum} 
The key idea in understanding the pendulum is the requirement that
the motion is determined by constraining forces in the
$\widetilde{x},\widetilde{y},\widetilde{z}$ frame. In the specific
case of pendulum motion, the constraint is that the pendulum bob
move in a circular arc. We will, however, not immediately confine
ourselves to the case of a circular-motion pendulum, but will keep
the description as general as possible for as long as possible.
Initially we will suppose only that the motion can be described by
two functions $\widetilde{x}(\tau)$,\,
$\widetilde{y}(\tau)$.\cite{3D} 
The
constraint on the motion (for example, that the particle move in an
arc of radius
$L$) can be thought of a curve in the
$\widetilde{x}$-$\widetilde{y}$ plane.

We are faced now with the task of combining a description of a
constrained path in the earth-like
$\widetilde{x},\widetilde{y},\widetilde{z}$ frame, with an
understanding of gravity (there is no gravity) in an inertial
frame. 
To do this, we consider a single
moment for the pendulum bob, and invoke the $t',x',y',z'$ Minkowski
system that is instantaneously comoving with the earth-like
frame.\cite{comoving} Because the earth-like frame and the
instantaneously comoving frame are momentarily at rest with respect
to each other, the spatial directions at that moment are the same in
the two frames. Then in the instantaneously comoving frame, we can
describe both the constraining path and the (nonexistent) nature of
gravity.

We now let 
\begin{equation}
U^{t'}\equiv\frac{dt'}{d\tau},\quad
U^{x'}\equiv\frac{dx'}{d\tau},\quad
U^{y'}\equiv\frac{dy'}{d\tau}
\end{equation}
be the primed-frame components of the 4-velocity of the particle,
so that the components of the acceleration 4-vector
are
\begin{equation}
a^{t'}\equiv\frac{d U^{t'}}{d\tau}=\frac{d^2t'}{d\tau^2},\quad
a^{x'}\equiv\frac{d U^{x'}}{d\tau}=\frac{d^2x'}{d\tau^2},\quad
a^{y'}\equiv\frac{d U^{y'}}{d\tau}=\frac{d^2y'}{d\tau^2}\,.
\end{equation}
The requirement that there is no acceleration in the direction of
motion means that 
\begin{equation}\label{notangent} 
a^{x'}U^{x'}+a^{y'}U^{y'}=0\ .
\end{equation}
For any motion,\cite{aperpU} the 4-acceleration and the 4-velocity
satisfy
\begin{equation}\label{adotU}
c^2
a^{t'}U^{t'}-a^{x'}U^{x'}-a^{y'}U^{y'}=0\ .
\end{equation}
Equations (\ref{notangent}) and (\ref{adotU})
tell us that
$a^{t'}U^{t'}$ must be zero, and hence that $a^{t'}$ must be zero.
(The $U^{t'}$ component cannot be zero.)

We take this result
as the key to the
dynamics:
$a^{t'}=0$ in the inertial frame that is momentarily comoving with
the earth frame. If the momentary $x$ velocity of the
instantaneously comoving ($t',x',y',z'$) frame\cite{newbeta} is
$c\beta$ with respect to the
$x,y,z,t$ frame, then the Lorentz transformation tells us that
\begin{equation}
\label{notangent2} 
a^{t'}=\frac{1}{\sqrt{1-\beta^2}} \big(
a^{t}-\beta a^{x}/c
\big) .
\end{equation}
It is straightforward to show (and is explicitly shown in
Appendix~\ref{sec:appendixA}) that the Minkowski frame comoving
with the rocket at event $t,x$ has $\beta=ct/x$.
The condition in Eq.~(\ref{notangent2}) for 
no acceleration along the motion becomes
\begin{eqnarray}\label{accel} 
0&=&a^{t}-\beta a^{x}/c 
=\frac{d^2t}{d\tau^2}-\frac{t}{x}\frac{d^2x}{d\tau^2}
\nonumber \\
 &=&\frac{1}{x}\;\frac{d}{d\tau}
\Big[x^2\frac{d}{d\tau}
\big(\frac{t}{x}
\big)
\Big]. \label{14}
\end{eqnarray}
{}From Eq.~(\ref{14}) we infer 
\begin{equation}
x^2\,\frac{d}{d\tau}\big(\frac{t}{x}\big)
=x\frac{dt}{d\tau}-t\frac{dx}{d\tau}=\mbox{constant}\equiv K,
\label{15}
\end{equation}
and from Eq.~(\ref{SRTxoft}) we have
\begin{equation}
\label{16}
x\frac{dx}{d\tau}-c^2t\frac{dt}{d\tau}
=\big(\widetilde{x}+{c^2}/{g}\big) \frac{d\widetilde{x}}{d\tau}.
\end{equation}
From Eqs.~(\ref{15}) and (\ref{16}) and from Eq.~(\ref{SRTxoft}),
we can solve for $dx/d\tau$ and $dt/d\tau$ in terms of
$d\widetilde{x}/d\tau$:
\begin{eqnarray}
\frac{dx}{d\tau}&=&\frac{c^2tK}{(\widetilde{x}+{c^2}/{g})^2}
+\frac{x}{(\widetilde{x}+{c^2}/{g})}\;
\frac{d\widetilde{x}}{d\tau}\label{dxdtau} \\
\frac{dt}{d\tau}&=&\frac{xK}{(\widetilde{x}+{c^2}/{g})^2}
+\frac{t}{(\widetilde{x}+{c^2}/{g})}\;
\frac{d\widetilde{x}}{d\tau}\label{dtdtau}.
\end{eqnarray}

The differential of proper time $d\tau$ along the worldline of the 
pendulum bob, in terms of differentials of the inertial coordinates,
is
\begin{equation}
\label{dtau2} 
(d\tau)^2
=(dt)^2-c^{-2}(dx)^2-c^{-2}(dy)^2.
\end{equation}
We now substitute $dy=d\widetilde{y}$ and the
results in Eqs.~(\ref{dxdtau}) and (\ref{dtdtau}) into
Eq.~(\ref{dtau2}) to arrive at an expression for the motion
entirely in terms of
$\widetilde{x}(\tau)$ and $\widetilde{y}(\tau)$:
\begin{equation}
\label{genresult} 
\Big(\widetilde{x}+\frac{c^2}{g}\Big)\sqrt{c^2
+\Big(\frac{d\widetilde{x}}{d\tau}\Big)^2
+\Big(\frac{d\widetilde{y}}{d\tau}\Big)^2\;}=\mbox{constant} .
\end{equation}
\begin{figure}[h]
 \begin{center}
{\epsfxsize=90pt \epsfbox{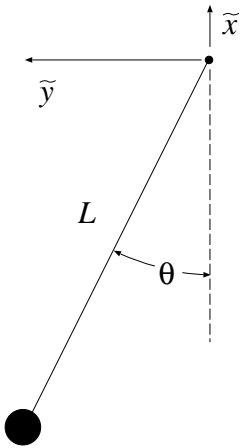}}
 \end{center}
\caption{\label{fig:pendulum} A pendulum making an arc of a circle in
the $\widetilde{x},\widetilde{y}$ plane. }
\end{figure}

Equation~(\ref{genresult}) is a single equation for the two
functions
$\widetilde{x}(\tau),\, \widetilde{y}(\tau)$.
This equation gives us the basis for describing
whatever constrained motion we wish, once we have chosen the path
-- the relationship between
$\widetilde{x}$ and
$\widetilde{y}$ imposed by the constraint. There can certainly be
different opinions about what constitutes the path of a pendulum bob
in relativity, but the choice made here is the most obvious. As
pictured in Fig.~\ref{fig:pendulum}, the pendulum bob maintains a
distance $L$ from the pivot, as measured in the
$\widetilde{x},\widetilde{y}, \widetilde{z}$ frame. 
With the pivot at the
$\widetilde{x}$, $\widetilde{y}$ origin, the constraint is that
$\widetilde{x}^2+\widetilde{y}^2=L^2$. This constraint can be
written as
\begin{equation}
\label{thetaeq} 
\widetilde{x}=-L\cos\theta(\tau),
\qquad \widetilde{y}=L\sin\theta(\tau)\,,
\end{equation}
where $\theta$ is the angle shown in Fig.~\ref{fig:pendulum}.\cite{realtheta} 
In terms of $\theta(\tau)$, Eq.~(\ref{genresult}) takes the 
form
\begin{equation}
\Big(-L\cos\theta+\frac{c^2}{g}\Big)\sqrt{c^2
+L^2\Big(\frac{d\theta}{d\tau}\Big)^2}=\mbox{constant}=
c\Big(-L\cos\theta_{\max}
+\frac{c^2}{g}\Big),
\end{equation}
where we have defined $\theta_{\max}$ as the maximum angular
excursion of the pendulum, the angle at which $d\theta/d\tau=0$.

We can now solve for $d\tau/d\theta$ and integrate to find the length
of proper time for a quarter of a period $P$,
\begin{equation}
\label{period} 
\frac{P}{4}=\int_0^{\theta_{\max}}
\Big(\frac{d\tau}{d\theta}\Big) d\theta
=\frac{L}{c}\!\int_0^{\theta_{\max}}
\Big[\Big(
\frac{-L\cos\theta_{\max}
+c^2/g}{-L\cos\theta+c^2/g}
\Big)^2
-1
\Big]^{-1/2}
d\theta.
\end{equation}
The value of $P$ given by Eq.~(\ref{period}) is smaller than 
the standard small angle period 
\begin{equation}
P_0=2\pi\sqrt{\frac{L}{g}} .
\end{equation}

\begin{figure}[h]
 \begin{center}
{\epsfxsize=160pt \epsfbox{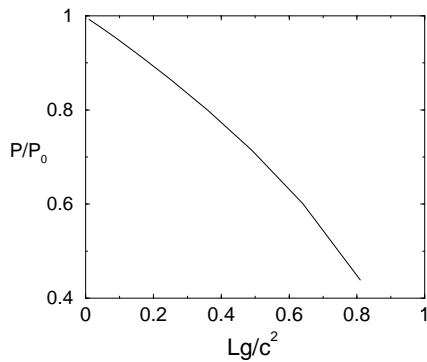}}
 \end{center}
\caption{\label{fig:period} 
The reduction of the period as a function 
of ${Lg/c^2}$ for $\theta_{\rm max}=5^\circ$.}
\end{figure}

In Fig.~\ref{fig:period}, the ratio $P/P_0 $ is plotted
as a function of $Lg/c^2$ for the case $\theta_{\max}=5^\circ$.
For extremely long $L$, comparable to $c^2/g\approx10^{16}$\,m, the
reduction is very significant. And this reduction cannot be
ascribed simply to the slowing of proper time for a rapidly moving
object. For example, the maximum value of $v\equiv Ld\theta/d\tau$
is 
$\approx0.175\,c$ for $\theta_{\max}=5^\circ$ and
$Lg/c^2=0.8$, corresponding to a time dilation factor of
$\sqrt{1-v^2/c^2}
\approx 0.985$. The reduction shown in 
Fig.~\ref{fig:period} is much greater than this.

\section{Conclusions}\label{sec:conclusions} 
``Special relativistic gravity'' is a fictitious force
arising in a noninertial earth-like reference frame. 
We have shown that ballistic and pendulum motions can be analyzed 
in this frame by using the principle that there is no gravity 
in a freely falling reference frame. This analysis makes good
pedagogical exercises, though of considerably different difficulty.
The study of ballistic motion is simple, while that of pendulum
motion brings in more physical ideas and somewhat trickier
mathematics.

\section{Acknowledgments}
This work was started at the University of Utah and
supported by the National Science Foundation under
grants PHY9734871 and PHY0244605. I thank an anonymous referee for
encouraging a reworking of the pendulum analysis; an earlier version
of the manuscript gave only the more advanced presentation now
relegated to Appendix~\ref{sec:appendixB}.

\appendix

\section{Calculations with geodesics and covariant differentiation
}\label{sec:appendixB} 

Here we exploit more advanced mathematical techniques to simplify
the calculations we discussed in the main text. For simplicity
we use units in which
$c=1$. We can describe the relationship of the inertial
$x,y,z$ system to the earth-based
$\widetilde{x},\widetilde{y},\widetilde{z}$ system with the
transformation
\begin{subequations}
\begin{eqnarray}
t&=&(\widetilde{x} +1/g)\sinh{g\widetilde{t}}\\
x&=&(\widetilde{x}+1/g)\cosh{g\widetilde{t}}\\
y&=&\widetilde{y}\\
z&=&\widetilde{z}.
\end{eqnarray}
\end{subequations}
In the noninertial
$\widetilde{x},\widetilde{y},\widetilde{z},\widetilde{t}$ coordinates,
the metric takes the form known as the Rindler
geometry\cite{rindler,rigidmetric}
\begin{equation}\label{rindmet} 
ds^2=-(1+g\widetilde{x})^2\,dt^2+d\widetilde{x}^2
+d\widetilde{y}^2+d\widetilde{z}^2. 
\end{equation}

A ballistic trajectory is a geodesic worldline, so 4-velocity
components must satisfy the geodesic equation
$DU^\alpha/d\tau=0$. From this it is easy to show that
$U^{\widetilde{y}}$ and $U_{\widetilde{t}}$ are constant, and we
choose the constants to be denoted by $u$ and $E$ respectively. The
fact that
$\vec{U}\cdot\vec{U}=-1$ implies that
\begin{equation}
1=\frac{(U_0)^2}{
(1+g\widetilde{x})^2
}-(U^{\widetilde{y}})^2-(U^{\widetilde{x}})^2
=\frac{E^2}{
(1+g\widetilde{x})^2}- u^2-\Big(\frac{d\widetilde{x}}{d\tau}\Big)^2.
\end{equation}
If we choose $\widetilde{x}$ to be zero when ${d\widetilde{x}}/{d\tau}
$ is zero, then $E^2=1+u^2$. The resulting differential equation is
\begin{equation}
\big(\frac{d\widetilde{x}}{d\tau}\big)^2=(1+u^2)
\Big[\frac{1}{
(1+g\widetilde{x})^2}-1\Big] .
\end{equation}
It is easy to check that $\widetilde{x}(\tau)$ 
given by Eq.~(\ref{traj}) is the solution to this differential
equation for $\widetilde{x}=0$ when $\tau=0$.

For the motion of the pendulum bob, we use the fact that in the
``stationary''
$\widetilde{x},\widetilde{y},\widetilde{z},\widetilde{t}$,
coordinates, the acceleration of the bob must have no component
along the motion, or
\begin{equation}
a^{\widetilde{x}}U^{\widetilde{x}}+a^{\widetilde{y}}U^{\widetilde{y}}=0\ .
\end{equation}
Because $\vec{a}\cdot\vec{U}=0$, it follows that
$a^{\widetilde{t}}=0$, or, equivalently,
$a_{\widetilde{t}}=0$. Using covariant differentiation in the 
$\widetilde{x},\widetilde{y},\widetilde{z},\widetilde{t}$,
coordinates, we have that
\begin{equation}
a_{\widetilde{t}}=\frac{dU_{\widetilde{t}}}{d\tau}
-U_{\widetilde{t}}\;U^{\widetilde{x}}\;
\Gamma^{\widetilde{t}}_{\widetilde{t}\widetilde{x}}
-U_{\widetilde{x}}\;U^{\widetilde{t}}\;
\Gamma^{\widetilde{x}}_{\widetilde{t}\widetilde{t}}=0 .
\end{equation}
It is straightforward to check that the Christoffel terms cancel 
each other, so we can conclude that $U_{\widetilde{t}}$ is 
a constant.\cite{conservedU} From $\vec{U}\cdot\vec{U}=-
1$, we have
then that 
\begin{equation}
1=\frac{(U_{\widetilde{t}})^2}{
(1+g\widetilde{x})^2
}-(U^{\widetilde{y}})^2 (U^{\widetilde{x}})^2,
\end{equation}
or 
\begin{equation}
(1+ g\widetilde{x})^2
\big[1+\big(\frac{d\widetilde{x}}{d\tau}\big)^2
+\big(\frac{d\widetilde{y}}{d\tau}\big)^2 \big]={\rm constant} ,
\end{equation}
which is identical to Eq.~(\ref{genresult}).

\section{The three special properties of the earth-like system
}\label{sec:appendixA} 

Brief derivations are given here of the three properties stated
in Sec.~\ref{sec:earthlike}.
The derivations will make use of the Minkowski coordinate system
$t',x',y',z'$ in the reference frame that is comoving with
rocket 1 at the point $P_1$ (coordinates $t_1,x_1$) on the
rocket's worldline, as shown in Fig.~\ref{fig:spacetime}. The idea
is that $\widetilde{x},\widetilde{y},\widetilde{z}$ are not part of
a Minkowski coordinate system, so we cannot directly apply to it
simple Lorentz transformations. But we {\em can} apply the
mathematics of Minkowski systems to $t',x',y',z'$.

For convenience we will hide some bothersome factors of $c$ by
introducing the common notation of a 0 coordinate $x^0\equiv ct$,
and a 0 component of 4-vectors, such as $U^0\equiv cU^t$ for the 
time component of the 4-velocity.

\subsection*{Simultaneity of rocket speed}
We first show that any rocket ``sees'' all other rockets to be at
rest with respect to itself. More specifically, we will show that
at a given moment of $t'$, all rockets have the same speed with
respect to the $t,x,y,z$ Minkowski reference frame (and hence with
respect to any Minkowski reference frame). We start by noticing
that in Fig.~\ref{fig:spacetime} both 
$P_1$ and $P_2$ lie on 
the $x'$ axis, the set of 
events with the same value of $t'$. The equation for that axis is
$dt'=0$. From the Lorentz transformation between the primed and
unprimed system this condition gives us 
\begin{equation}
\label{slope1} 
cdt-\beta dx=0,
\end{equation}
as the equation for the axis, where $c\beta$ 
is the speed of the primed frame with respect to the unprimed
frame. The value of $c\beta$ is simply $dx/dt$ for worldline 1 at
point $P_1$. From Eq.~(\ref{SRTxoft}), $c\beta=dx/dt=c^2t_1/x_1$,
where $t_1,x_1$ are the coordinates of $P_1$. We can now combine
this result with Eq.~(\ref{slope1}) to find that the slope 
of the $x'$ axis is
\begin{equation}\label{slope2} 
\frac{dt}{dx}=\frac{t_1}{x_1}\,. 
\end{equation}
Because the axis must go through the point $t_1,x_1$, it follows
that the equation of the axis is 
\begin{equation}
\frac{t}{x}=\frac{t_1}{x_1}\,. 
\end{equation}
The $x'$ axis is then simply the line going through the $t,x$
origin and through $P_1$.

Clearly we would have obtained precisely the same line if we had
started with point $P_2$, thus the $x'$ axis would be the same for
the Minkowski coordinate system comoving with the rocket at point
$P_2$. 
But if the $x'$ axis is the same, then
the reference frame comoving with rocket 1 at $P_1$ is the same as
the reference frame comoving with rocket 2 at $P_2$. In other words,
the speed of rocket 1 at $P_1$ is the same as that of rocket 2 at
the same moment of comoving time.

\subsection*{Rocket acceleration}
From Eq.~(\ref{SRTxoft}) we have that along a rocket worldline
$dx/dt=c^2t/x$, and hence the Lorentz factor is
\begin{equation}
\gamma=1/\sqrt{1-(dx/cdt)^2\;}=\frac{x}{\widetilde{x}+c^2/g}.
\end{equation}
The components of the 4-velocity and 4-acceleration are
\begin{equation}
U^0=c\frac{dt}{d\tau}=c\gamma=\frac{cx}{\widetilde{x}+c^2/g}
\quad\quad\quad
U^x=\frac{dt}{d\tau}\,\frac{dx}{dt}=\frac{c^2t}{\widetilde{x}+c^2/g}\,,
\end{equation}
and 
\begin{equation}
a^0=\gamma\;\frac{cdx/dt}{\widetilde{x}+c^2/g}
=\frac{c^3t}{(\widetilde{x}+c^2/g)^2},
\qquad
a^x=\gamma\;\frac{c^2}{\widetilde{x}+c^2/g}=
\frac{c^2x}{(\widetilde{x}+c^2/g)^2}\,.
\end{equation}

The quantity $a\equiv\sqrt{\vec{a}\cdot\vec{a}\;}$ is an 
invariant that signifies the acceleration ``felt'' by 
each rocket. (It is, for example, the component $a^{x'}$
of the acceleration, when evaluated in an instantaneously 
comoving Minkowski reference frame.)
With the above results we can evaluate $a$
to be
\begin{equation}
\sqrt{(a^x)^2-(a^0)^2
\;}=\frac{c^2}{\widetilde{x}+c^2/g}\,.
\end{equation}
The scalar $a$ is then constant along the worldline of each rocket,
but varies slightly from rocket to rocket. (For $\widetilde{x}$
small compared to $10^{16}$\,m, the variation in $a$ is
negligible.)

\subsection*{Rigidity of the earth-like frame}

The third important property of the earth-like reference frame
is its spatial rigidity, the time independence of the separation
of the rockets. More precisely, this property is that in a
reference frame instantaneously comoving with rocket 1, the
distance measured to rocket 2 will be the same at all times; it
will not depend on our choice of point $P_1$ on the worldline.

To prove this we start with the distance as measured in the comoving
frame at $P_1$. This measurement is simply $x_2'-x_1'$
made at a single moment of $t'$. It can be written in 
the form
\begin{equation}
x_2'-x_1'=\sqrt{(x_2-x_1)^2-c^2(t_2-t_1)^2}\,.
\end{equation}
Now $ct_1=\beta x_1$ and $ct_2=\beta x_2$ where $c\beta$
is the speed with respect to the $t,x,y,z$ system of rocket 1
at point $P_1$ or of rocket 2 at $P_2$. (It was shown above that
they are the same.) It follows that the distance is
\begin{equation}
\sqrt{(x_2-x_1)^2-c^2(t_2-t_1)^2\;}
=\gamma^{-1}(x_2-x_1) .
\end{equation}
We can next use $x_2=\gamma(\widetilde{x_2}+c^2/g)$
and
$x_1=\gamma(\widetilde{x_1}+c^2/g)$ 
to write the result as 
\begin{equation}
\mbox{distance}=\widetilde{x_2}-\widetilde{x_1}\,.
\end{equation}
This completes the proof that the distances separating rockets are
constant in time.


\begin{thebibliography}{33}
\bibitem{different_g} The notation here is slightly different
from that used by Desloge.\cite{rindlerrefs} The symbol $g$
used there, and that used here are related by
$g_{\rm desloge}={c^2}/({\widetilde{x}+c^2/g_{\rm here}})$.
\bibitem{explainrindler} The range of $x,t$ to which
Eq.~(\ref{SRTxoft}) applies is limited to $|x|\geq ct$, but there
are no real boundaries to the spacetime at $x=\pm ct$. These are
boundaries only of the region that are covered by
$\widetilde{x}$ values in Eq.~(\ref{SRTxoft}).
\bibitem{3D} It is only for convenience that we are assuming that
the motion is planar. The analysis to follow is very easily
extended to three-dimensional motion by adding a
$(dz)^2$ term to Eq.~(\ref{dtau2}).
\bibitem{comoving} Note that the comoving frame is not 
instantaneously comoving with the pendulum bob. Rather, at the
moment in question it is comoving with the rocket at the location
of the pendulum bob. By the special properties of the worldlines of
Eq.~(\ref{SRTxoft}) it also is instantaneously comoving with all
other rockets.
\bibitem{aperpU} The orthogonality of the 4-velocity and 4-acceleration is also
standard. See, for
example, Ref.~\onlinecite{mtw}, Eq.~(6.2). It follows immediately
from differentiating 
$c^2 U^{t'\,2}-U^{x'\,2} -U^{y'\,2}=c^2$ with respect to $\tau$.
\bibitem{newbeta} The velocity parameter $\beta$ here has no direct relation to
the velocity parameter $\beta$ used in Sec.~\ref{sec:ballistic}.
\bibitem{realtheta} The variable $\theta$ is illustrated by, but
not defined by, Fig.~\ref{fig:pendulum}. The definition is 
$\theta\equiv\tan^{-1}( {-\widetilde{y}/\widetilde{x}})$.
\bibitem{rigidmetric}The fact that the spatial part of this metric
is independent of time is related to the point made in
Sec.~\ref{sec:earthlike} that the spatial distances between rockets,
measured in an instantaneously comoving frame, do not change in
time.
\end{thebibliography}
\end{document}